\documentclass[11pt,a4paper]{article}

\usepackage{amsmath}	
\usepackage{braket}
\usepackage[margin=2.5cm]{geometry}
\begin{document}

\title{VSFT revisited}

\author{Syoji Zeze\\
 Yokote Seiryo Gakuin High School, Yokote 013-0041, Japan}

\maketitle

\begin{abstract}%
We argue validity of purely ghost kinetic operator 
in open string field theory from the perspective of the modern 
analytic method based on the $K B c$ subalgebra.  A purely ghost 
kinetic operator is obtained as a result of gauge fixing
string field theory around the identity based tachyon vacuum solution. 
It is shown that the obtained kinetic operator is not equivalent to the
midpoint insertion of the conformal ghost which is
extensively studied in literature.  We also find that
the equation of motion does not allow any nontrivial solutions.  
\end{abstract}

\section{Introduction} 

The discovery of the analytic solution in open string field
theory \cite{Schnabl:2005gv} and subsequent developments
have provided a cornerstone to understand the nonperturvative 
nature of string field theory.  Recently, it has been recognized 
that multiple D-branes solutions \cite{Erler:2011tc, 
Murata:2011ex,
Murata:2011ep, Hata:2011ke, Hata:2012cy, Masuda:2012kt} 
suffer from an anomaly.  It reflects 
nontrivial nature of the space of the string
field, which is not yet fully understood. 
With such a hard problem, it might be helpful
to understand the underlying structure of the space of D-brane solutions
\footnote{Hata and Kojita 
have identified a number of D-branes as a winding number \cite{Hata:2011ke,
Hata:2012cy}, while their solutions
are also singular for large winding number. }.
It would be worth recalling that vacuum string field theory (VSFT) 
\cite{Rastelli:2001uv}, 
whose kinetic operator is purely ghost \cite{Horowitz:1987kz,
Rastelli:2000hv}, provides very simple description of D-branes. 
The equation of motion can be factorized into matter and ghost parts,
and the matter part obeys the projector equation $\Psi_{m}^2
= \Psi_m$ \cite{Rastelli:2001jb}. 
With the help of the simplicity, many important works
such as ratio of the D-brane tension \cite{Rastelli:2000hv,
Rastelli:2001jb}, 
surface states as projectors \cite{Rastelli:2001jb, Rastelli:2001vb},
and relevance to noncommutative field theory \cite{Dasgupta:2000ft,
Gopakumar:2000zd, Witten:2000nz, Chen:2002jd} had been done.  If such
simple prescription remains valid and can be realized in modern analytic
context, it will be helpful to understand the nature of D-branes in
string field theory.

Historically, VSFT was conjectured as a candidate for the tachyon 
vacuum before the analytic description became available.  Now
the analytic and closed expressions are already available \cite{Schnabl:2005gv,
Okawa:2006vm, Erler:2009uj} so we can apply them to examine whether 
VSFT is valid description of string field theory around the tachyon vacuum.
At first look, a kinetic operator derived from an analytic solution
contains matter pieces hence contradicts those of VSFT.
On the other hand, some aspects of the analytic
solutions seem to be relevant to VSFT.  
One of them is the appearance of a sliver-like phantom term in some analytic solutions,
which implies underlying projector-like structure \cite{Schnabl:2005gv,
Ellwood:2009zf, Erler:2012qr, Erler:2012qn}.  
The other aspect, which will be studied in this letter,
is existence of identity based solutions whose kinetic operator
is very close to but not exactly pure ghost \cite{Erler:2009uj,
Erler:2012qn, Zeze:2010sr, Inatomi:2011xr}. In this letter, we  
derive VSFT from the SFT expanded around an
identity based solutions $c - c K$.  Our results are summarized as below.
\begin{itemize}
 \item A purely ghost kinetic operator can be obtained by gauge 
       fixing SFT around an identity based solution.  No singular
       reparametrization \cite{Gaiotto:2001ji} is required. 
 \item The kinetic operator is given by an insertion of conformal ghost
       at the boundary of open sting world sheet, hence is not
       equivalent to the midpoint insertion studied in past \cite{Gaiotto:2001ji}. 
 \item There are no nontrivial classical solution.  This is due to
       the gauge condition imposed on the string field. 
\end{itemize}
Contrary to our motivation,  these results
indicate that the VSFT derived from the analytic solution is inconsistent.

\section{Gauge invariant action}

The VSFT conjecture \cite{Rastelli:2001uv} claims that a pure ghost
kinetic operator can be obtained from string field theory expanded around
the tachyon vacuum solution even before gauge fixing.  
We would like to reexamine the conjecture with the help of the analytic
solutions based on the $K B c$ technique \cite{Okawa:2006vm}.   
We first assume that the tachyon vacuum solution $\Psi_{V}$ that realizes VSFT
is gauge equivalent to an analytic solution $\Psi_{O}$ written in Okawa
form $F c K (1-F^2)^{-1} B c F$ \cite{Okawa:2006vm} or its real form \cite{Erler:2009uj}.
Therefore we can write
\begin{equation}
 \Psi_{V} =U^{-1} Q_{B} U + U^{-1} \Psi_{O} U,
\end{equation}
where $U$ is a gauge element.  The VSFT action should be obtained from the
SFT action expanded around $\Psi_V$.  The action can be
evaluated as
\begin{align}
S_{Q_B}[\Psi+ \Psi_{V}] &= S_{Q_B}[\Psi+ U^{-1} Q_{B} U + U^{-1}
 \Psi_{O} U] \label{actions} \\  
  & = S_{Q_B}[U^{-1} Q_B U
+ U^{-1}    (   
\Psi_{O} + U \Psi U^{-1})U
] \notag \\ 
 & = S_{Q_B}[\Psi_{O}   + U \Psi U^{-1}] \notag \\ 
 & = S_{Q_{O}}[ U \Psi U^{-1}], \notag
\end{align}
where $S_{Q_{B}}$ and $S_{Q_{O}}$ are SFT actions whose kinetic
operators are $Q_{B}$ (the usual BRST operator) and $Q_{O}$ (the kinetic
operator derived from $\Psi_O$), respectively.  
We have used the gauge invariance of the SFT action between second and
third lines in (\ref{actions}).   The last line of (\ref{actions})
means that the VSFT action is equivalent to the action defined
by the kinetic operator $Q_{O}$ up to field redefinition.  Therefore we
next discuss whether the kinetic operator $Q_{O}$ can be pure ghost. 
The expression of the kinetic operator $Q_{O}$ is given by
\begin{equation}
 Q_{O} \Psi =   Q_{B} \Psi +  \Psi_{O} \Psi+\Psi_{O} \Psi.   \label{125705_19Sep13}
\end{equation}
Since matter part contributions in (\ref{125705_19Sep13}) only appear 
through the matter Virasoro generators $L^{X}_{n}$, $Q_O$ should satisfy  
$\delta Q_{O}/ \delta{L^{X}_n}=0$ for each $n$ if it is pure ghost.
This leads to the following condition
\begin{equation}
c_{n} \Psi  +  \frac{\delta \Psi_O}{\delta L^{X}_{n} }  \Psi + \Psi \frac{\delta \Psi_O}{\delta L^{X}_{n} }
 =0,\label{130705_19Sep13} 
\end{equation}
where $c_{n}$ is a mode of the conformal ghost $c(z)$.  
In order to satisfy (\ref{130705_19Sep13}),  $\Psi_{O}$ should be linear in $L_{n}^{X}$ hence $K$.
This arrows only few candidates from the variety of $F$ in $\Psi_{O}$,
i.e.,  the  identity based solutions discussed in \cite{Erler:2012dz}.   
However, it is easily understood that they 
do not satisfy (\ref{130705_19Sep13}) 
since they involve $K$
only through terms such as $ c K $, $K c$ or $c K B c$, and variations of
these terms with respect to $L^{X}_n$ are proportional to $c \sim c(1)$,
not $c_n$.  Therefore, there is no chance to obtain pure ghost kinetic
operator in gauge invariant action 
under the assumption that the tachyon vacuum is gauge equivalent
to Okawa's solution.  

\section{Purely ghost kinetic operator in gauge fixed theory}

While pure ghost kinetic operator can not be realized in the gauge
invariant action, there still be a chance to obtain pure ghost 
kinetic term in gauge fixed theory.  Most suitable candidate 
is the identity base solution 
\begin{equation}
 \Psi_{0} = c-c K\label{005419_26May13}.
\end{equation}
While it has a drawback of being singular without regularization, 
it still has been believed to be a consistent description of the  
tachyon vacuum \cite{Zeze:2010sr, Erler:2012dz}. 
The kinetic operator around this solution is defined by
\begin{align}
 \mathcal{Q} \Psi &= Q_B \Psi + \Psi_{0} \Psi 
-(-1)^{\Psi}\Psi \Psi_{0}  \notag \\ 
& = Q_{B} \Psi + c (1-K )\Psi -(-1)^\Psi
\Psi c (1-K).  
\end{align}
Obviously, the kinetic operator $\mathcal{Q}$ is not pure ghost due to the $c K$
pieces and the original BRST charge.  However, it is quite suggestive
that the kinetic operator acts on $c$ and $\mathcal{B}$ as
\begin{equation}
\mathcal{Q} c =0, \qquad \mathcal{Q} B = 1, \label{083844_17May13}
\end{equation}
as if $\mathcal{Q}$ is replaced by an adjoint action of
purely ghost operator $\{c,*\}$.  
In addition, the latter equation of (\ref{083844_17May13}) indicates
that the homotopy operator of $\mathcal{Q}$, which ensures trivial
cohomology at the tachyon vacuum, is just given by $B$.
These facts tempt us to identify  $\mathcal{Q}$ as $ \{c, *\}$.
To realize this identification, it should be noticed that
equations (\ref{083844_17May13}) do not define $\mathcal{Q}$ uniquely.  
For example, a replacement $\mathcal{Q} \rightarrow \mathcal{Q} + \{c',
*\}$ does not alter the
equations as long as $c'$ anticommutes with both $c$ and $B$.  It is
expected that such redundant components of $\mathcal{Q}$ can be removed
by suitable gauge fixing.  Suppose that we find a set of
operators which satisfies
\begin{equation}
 \{\mathcal{Q} , \mathcal{B} \} = 1, \quad \{\mathcal{C} , \mathcal{B} \} = 1,\label{131140_17May13}
\end{equation}
where $\mathcal{C}$ and $\mathcal{B}$ are linear
combinations of $c_n$ and $b_n$ respectively. We also require 
$\mathcal{B}$ to be Hermite.  Then we impose  `linear $B$ gauge'
condition \cite{Kiermaier:2007jg} on the string field as
\begin{equation}
 \mathcal{B} \Psi =0.
\end{equation} 
Then, inserting $\{\mathcal{C} , \mathcal{B}\}=1$
in the kinetic term,  we arrive at a purely ghost kinetic operator
\begin{equation}
 \mathrm{Tr} [\Psi \mathcal{Q} \Psi] = \mathrm{Tr} [\Psi \mathcal{C} \Psi], 
\end{equation}
as desired.  A solution of (\ref{131140_17May13}) is easily found 
and is expressed in the $K B c$ language as
\footnote{The gauge condition  $\mathcal{B}\Psi=0$ 
is equivalent to  $(\mathcal{B}_{0}+\mathcal{B}_{0}^{\dagger})\Psi 
=0$, in which perturbation theory is first discussed in \cite{Fuji:2006me}
and further elaborated in \cite{Kiermaier:2007jg}.}
\begin{equation}
 \mathcal{B} \Psi = \frac{1}{2}
\left\{B \Psi +(-1)^\Psi \Psi B\right\}, \quad
\mathcal{C} \Psi =  
c  \Psi   - (-1)^\Psi \Psi c.\label{212958_25May13}
\end{equation}

\section{Comparison with the VSFT conjecture}

While the kinetic operator we found is indeed pure ghost, there is a
significant difference between our result and the earlier version of VSFT
\cite{Erler:2012dz}.  Namely, our action is gauge fixed but the latter
is not.  Therefore, in principle, these two results can not be compared 
on an equal footing. However, it also should be remembered that the VSFT
conjecture was made when the details of the tachyon vacuum solution
is not available.  There still be a possibility such that, 
VSFT predicts the kinetic term of gauge fixed theory correctly while
the interpretation of the space of string fields is wrong. 
If so, earlier results derived from VSFT should be reconsidered 
according to suitable gauge condition consistent with the kinetic operator. 
In order to address such possibility, it is convenient to rewrite $\mathcal{C}$ in operator formalism
as $ c(1) - c(-1) = 2 \sum_{n=0}^{\infty} c_{2n}
$. This linear combination indeed
belongs to the general class of pure ghost kinetic operator
suggested in \cite{Rastelli:2000hv}.   
However, in the subsequent developments of VSFT, most 
attentions were payed to the specific choice of ghost insertion on the 
midpoint of an open string, i.e., $c(i) -c(-i)$, which does not coincide
with our result. We still have a chance to relate them
by a suitable conformal transformation.  
More explicitly, two kinetic operators can be
related by a conformal map $U$ such that
\begin{equation}
 U (c (1) - c(-1)) U^{-1} \sim c(i) -c (-i).
\end{equation}
where $U = \exp (i   \pi/2 \,  L_0 ) $, which gives a
rotation around the unit circle.

is not a symmetry of the gauge fixed theory since only those generated by
$K_n = L_{n} -(-1)^n L_{-n}$ leave gauge fixed action invariant.

Therefore, our result  $c(1) - c(-1)$, insertion of conformal ghosts on
the boundary, is not equivalent to $c(i) - c(-i)$ those on the midpoint 
even in gauge fixed theory. 

\section{Classical solutions}

Another implication comes from a study of the equation of motion.
In our prescription, the pure ghost kinetic operator is
obtained by gauge fixing.  Therefore, a gauge condition on the classical solution
is given from the beginning and can not be chosen arbitrary.
This situation is in contrast with those in most literature, where
Siegel gauge is chosen irrespective with the choice of the kinetic operators.
With this in mind, let us consider the equation of motion
\begin{equation}
 \mathcal{C} \Psi + \Psi^2 = 0 \label{154953_20May13}
\end{equation} 
where the string field obeys the gauge condition $\mathcal{B} \Psi =0$,
or equivalently $ B \Psi = \Psi B$ in the $K B c$ notation.  Multiplying
(\ref{154953_20May13}) by $\mathcal{B}$ from the left we have
\begin{equation}
 \Psi + 2 B \Psi^2 =0.  \label{155222_20May13}  
\end{equation}
Multiplying  (\ref{155222_20May13}) further by $B$ from the left yields
$B \Psi =0$, and plugging this back to (\ref{155222_20May13})
yields $\Psi=0$.   In this way, we arrive at the striking result 
that the equation of motion does not allow any nontrivial solution.
Obviously  this is pathological since we know that
the nontrivial solution $\Psi = - \Psi_0$ which represents original 
D25-brane is available before gauge fixing.  One may guess that
this pathology is due to the choice of too simplified kinetic
operator.  However, this is not the case. Consider more general
background than (\ref{005419_26May13}) according to Okawa \cite{Okawa:2006vm},
\begin{equation}
 \Psi_{O} = F c \frac{K}{1-F^2} B c F,\label{161453_3Jun13}
\end{equation}
where $F$ is a function of $K$.  In this case, 
the homotopy operator is given by $A= B (1-F^2)/K $. It should be noted
that we still have $A^2 = 0$ even though $A$ depends on $K$ nontrivially.   
A gauge condition analogous to (\ref{212958_25May13}) is
\begin{equation}
 \mathcal{A} \Psi = A \Psi +(-1)^{\Psi} \Psi A =0.
\end{equation}
Then, repeating same process as seen in
(\ref{154953_20May13}) and (\ref{155222_20May13}) for
the kinetic operator $\mathcal{Q}$ defined by (\ref{161453_3Jun13}), 
we obtain the trivial solution $\Psi=0$ 
again. Therefore, the gauge condition $\mathcal{A}\Psi=0$,
which looks honest with the trivial cohomology, does not allow
nontrivial solution.  Above augment shows that the origin of the pathology
is not due to too simplified kinetic term but due to too strong gauge
condition.

\section{Discussions}

We study a pure ghost kinetic operator in open string field theory by
gauge fixing a theory around identity based analytic solution.   The 
obtained pure ghost operator is placed on the boundary therefore is not
equivalent to the midpoint insertion.  It is also shown that the gauge
fixed equation of motion does not have any nontrivial solution. 
The latter result indicates that the VSFT obtained here cannot explain 
the `original' configuration which represents D25-brane.  
Above results indicate that the early implications form VSFT can not be 
applied directory to the string field theory around the analytic solutions.

While the gauge fixing examined in this letter fails to explain
D-branes, it may be still useful as a local description around
the tachyon vacuum.  For example, derivation of the effective action
around tachyon vacuum is possible along the line with
\cite{Rastelli:2000hv}.  Derivation of closed string amplitude 
\cite{Gaiotto:2001ji, Drukker:2002ct, Drukker:2003hh, Takahashi:2003kq}
will also  be interesting.  In either case, a regularization will be required
for vanishing area of the world sheet.

\section{Acknowledgements}

We thank H.~Isono and I.~Kishomoto for helpful discussions.

\end{document}